\documentclass[oneside,a4paper,reqno]{amsart}
\usepackage{amsmath}
\usepackage{amssymb}
\usepackage{xspace}
\usepackage{graphics}
\usepackage[mathscr]{eucal}

\theoremstyle{remark}

\newcommand{\bmat}[3]{\ensuremath{
																								\bigl \langle     #1   \bigr|    \, #2\,   \bigl|     #3   \bigr \rangle
                        									}
                     }

\newcommand{\ket}[1]{\ensuremath{		\left| #1 \right> 
																																			  }
																									}
\newcommand{\bket}[1]{\ensuremath{		\bigl| #1 \bigr> 
																																			  }
																									}

\newcommand{\overlap}[2]{\ensuremath{ 
																								\left \langle    #1 \vphantom{#2 } \,
                        \right| \left.   #2 \vphantom{#1}
                        \right \rangle
                        									}
                     }

\newcommand{\comm}[2]{\ensuremath{  \left[ #1, #2 \right] }}

\newcommand{\pt}{   \ensuremath{     \phi_{\mathrm{ap}}     }}
\newcommand{\xOp}{ \ensuremath{  \hat{x}  }}
\newcommand{\pOp}{ \ensuremath{  \hat{p}  }}

\newcommand{\MxOp}{  \ensuremath{   \hat{\mu}_{\mathrm{X}}    } }
\newcommand{\MpOp}{  \ensuremath{   \hat{\mu}_{\mathrm{P}}    } }

\DeclareMathOperator{\Det}{det}
\DeclareMathOperator{\sign}{sign}
\begin{document}
\begin{titlepage}
\begin{center}
\bfseries
RETRODICTIVELY OPTIMAL LOCALISATIONS IN PHASE SPACE
\end{center}
\vspace{1 cm}
\begin{center}
D M APPLEBY 
\end{center}
\begin{center}
Department of Physics, Queen Mary and
		Westfield College,  Mile End Rd, London E1 4NS, UK
 \end{center}
\vspace{0.15 cm}
\begin{center}
  (E-mail:  D.M.Appleby@qmw.ac.uk)
\end{center}
\vspace{1.35 cm}
\begin{center}
\textbf{Abstract}\\
\vspace{0.35 cm}
\parbox{10.5 cm }{  In a previous paper it was shown that the distribution of measured
                    values for a retrodictively optimal simultaneous measurement of 
                    position and momentum is always given by the initial state Husimi function.
                    This result is now generalised to retrodictively optimal simultaneous
                    measurements of an arbitrary 
                    pair of rotated quadratures $\xOp_{\theta_1}$ and $\pOp_{\theta_2}$.  It is
                    shown, that given any such measurement, it is possible to find another 
                    such measurement, informationally equivalent to the first, for which the
                    axes defined by the two quadratures are perpendicular.  It is further shown
                    that the distribution of measured values for such a meaurement belongs to 
                    the class of generalised Husimi functions most recently discussed 
                    by W\"{u}nsche and Bu\v{z}ek.  The class consists of 
                    the subset of W\'{o}dkiewicz's operational
                    probability distributions 
                    for which the filter reference state is a squeezed vaccuum state.
                 }\\     
\end{center}
\end{titlepage}
\section{Introduction}
\label{sec:  intro}
During the last few years there has been considerable progress in the problem of 
simultaneously measuring both the position and the momentum of a quantum
mechanical system\cite{Measure,LeonBook}.  In several recent
publications~\cite{self1,self3,self2} we have discussed  how to characterise 
the accuracy of, and disturbance caused by such measurements.  One approach to
the problem is that based on the concept of an ``unsharp
observable''~\cite{Davies,Fuzzy,Fuzzy2,Ban}.  This approach
has recently been criticised by Uffink~\cite{Uffink}.  In the papers just mentioned we
took a rather different
approach, based on methods developed by Braginsky and Khalili~\cite{Braginsky}.  It appears
to us that these methods have certain advantages, both conceptually (they clarify what
is meant by the term ``accuracy'' in a quantum mechanical context~\cite{self1}), and
practically (they facilitate the calculations~\cite{self3}).

Another advantage of these methods is that they give additional insight into
the physical significance of the Husimi function~\cite{Husimi,Reviews}.  The fact that the
Husimi function describes the distribution of measured values for many \emph{particular
cases} of joint measurement processes is, of course, well known~\cite{Measure,LeonBook}.  In
ref.~\cite{self2} (also see Prugove\v{c}ki and Ali~\cite{Fuzzy2}) we showed that the Husimi
function actually has a much stronger,
\emph{universal} property:  namely, it gives the distribution of results  for \emph{any}
retrodictively optimal measurement process (\emph{i.e.} any process which is
retrodictively unbiased, and which maximises the retrodictive accuracy).

The fact that the Husimi function gives the distribution of results whenever the measurement
is retrodictively optimal, and is otherwise independent of the details of the particular
process employed, could be interpreted to mean that the Husimi function plays the same
role for joint measurements of $x$ and $p$ that is played by the function
$|\overlap{x}{\psi}|^2$ for single measurements of $x$ only.

The purpose of this paper is to show that a similar universal property holds for
generalised Husimi functions of the form
\begin{align}
  Q_{\mathrm{gen}} (x,p)
& = \frac{1}{\pi} \int dx' dp' \,
     \exp \left[-\left( a (x - x')^2 + 2 c (x-x') (p-p') + b (p - p')^2 \right) \right]
\notag
\\
& \hspace{2.5 in} \times
     W(x',p')
\label{eq:  GenHusDefInt}
\end{align}
where $a$, $b$ and $c$ are real,  $a, b \ge 0$,
$ab-c^2 = 1$, and where $W$ is the Wigner function.
These functions are  operational distributions of the type defined by
W\'{o}dkiewicz~\cite{Davies,Ban,Wod,Lal1}.  They are the distributions which result
when the  filter reference state (or ``quantum ruler'') is an arbitrary squeezed
vacuum state~\cite{Squeeze}. They have  been discussed by  Halliwell~\cite{Halli},
W\"{u}nsche~\cite{Wuen} and W\"{u}nsche and Bu\v{z}ek~\cite{WuenBuz}.

\begin{figure}
\includegraphics{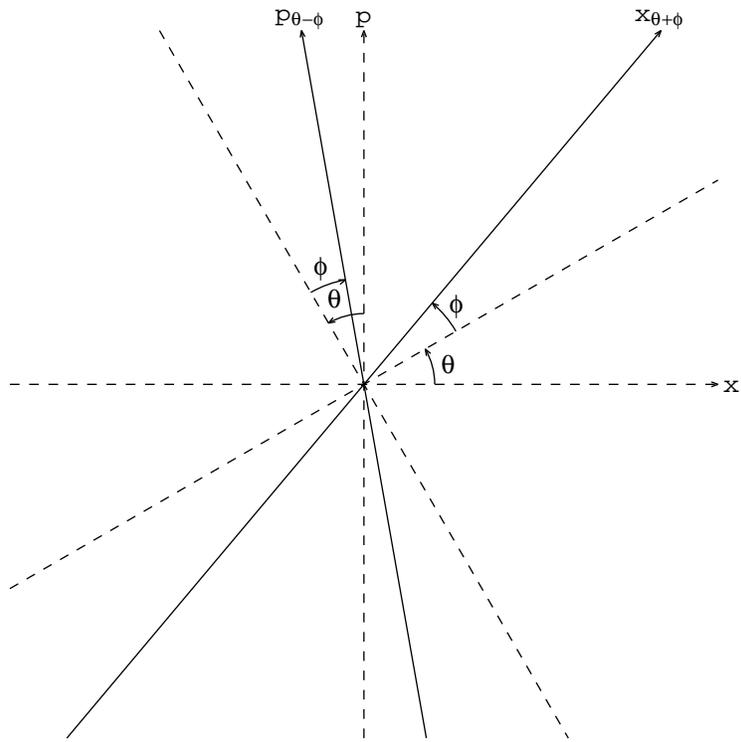}
\caption{Oblique axes:  parameterisation}
\label{fig:  oblique}
\end{figure}

In ref.~\cite{self2} we considered retrodictively optimal measurements of $x$ and $p$. 
However, this is clearly not the only way to determine the location of a
system in phase space.  What would happen if, instead of determining $x$ and $p$, one were to
make a retrodictively optimal measurement of an arbitrary pair of rotated quadratures, not
necessarily at $90^{\mathrm{o}}$ (see Fig.~\ref{fig:  oblique})?  Such measurements 
are  possible (using a suitably modified form of  homodyne
detection~\cite{LeonBook}, for example).  We will show that  the outcome of such a
measurement is always described by a generalised Husimi function of the kind defined by
Eq.~(\ref{eq:  GenHusDefInt}).  As with our previous result this is a universal statement: 
it only depends on the measurement being retrodictively optimal, 
and is otherwise independent
of the details of the particular process employed.

The main difficulty in proving this result comes from the fact that the 
measurements we consider are characterised by three independent parameters [namely,two
angles
$\theta,
\phi$ to specify the oblique coordinate system (see Fig.~\ref{fig:  oblique}), and a
parameter $\lambda$ to specify the relative accuracy of the measurements of the two
quadratures].  On the other hand it only needs two  parameters to specify a distribution of
the type defined by Eq.~(\ref{eq:  GenHusDefInt}).  It follows, that corresponding to any
given distribution, there is an infinite  set of informationally equivalent measurements. 
It is the  problem of characterising  these  sets, and giving a precise definition of
``informational equivalence,'' which will mainly concern us in the following.

One might also ask what is the significance of distributions which are like the ones
considered in this paper in that they are obtained from the Wigner function by smoothing it
with a Gaussian convolution, but in which the determinant $ab-c^2 >1$ (corresponding to an
impure filter reference state)~\cite{Lal1,Halli,Wuen,WuenBuz,Cart,GenHus}.  In
ref.~\cite{self3} we showed that, in the special case of the Arthurs-Kelly process, such
functions describe the outcome of measurements producing the smallest possible
amount of disturbance for a given, sub-optimal degree of accuracy.  It is natural to wonder
whether this property is also universal, and whether it also generalises to the case of
simultaneous measurements of an arbitrary pair of rotated quadratures.  However, that is a
question which we leave to the future.
\section{Linear Canonical Transformations}
\label{sec:  LinCanTrans}
Squeezed states arise as a result of making linear canonical transformations
of the creation and annihilation operators~\cite{Squeeze}.  We begin by describing the
parameterisation of these transformations which will be employed in the sequel.

Consider a system,
having one degree of freedom, with position
$\xOp$ and momentum
$\pOp$.  In some applications $\xOp$, $\pOp$ are dimensionless to begin with.  If not they
can be made dimensionless, by making the replacements $\xOp \rightarrow \frac{1}{l} \xOp$,
$\pOp \rightarrow \frac{l}{\hbar} \pOp$, where $l$ is a fixed, in general arbitrary constant
having the dimensions of length.  In the sequel we will always assume that this has been
done, so that $\comm{\xOp}{\pOp} = i$. 

We are interested in transformations of the form
\begin{equation*}
  \begin{pmatrix}
    \xOp_{M} \\ \pOp_{M}
  \end{pmatrix}
=
  M
  \begin{pmatrix}
    \xOp \\ \pOp
  \end{pmatrix}
\end{equation*}
where~\cite{Squeeze} $M$ 
 belongs to the group $\mathrm{SL}(2,\mathbb{R})$ [which is isomorphic to
$\mathrm{Sp}(2,\mathbb{R})$]. The group may be parameterised as
follows.  Given any matrix 
$M \in \mathrm{SL}(2,\mathbb{R})$  there exist
unique $\theta$ in the range $- \pi < \theta \le \pi$ and unique $\phi$ in the range
$-\frac{\pi}{4} < \phi < \frac{\pi}{4}$ such that
\begin{equation*}
  M 
=  \begin{pmatrix}
     \alpha \, \cos ( \theta + \phi ) &  \alpha \,\sin ( \theta + \phi ) \\
    - \beta \sin (\theta - \phi ) & \beta \cos ( \theta - \phi )
   \end{pmatrix}
\end{equation*}
for suitable positive constants $\alpha$, $\beta$ (see Fig.~\ref{fig:  oblique}).
The requirement that
$\Det M = 1$ means that $\alpha \beta \cos 2 \phi = 1$.  We may therefore write
\begin{equation}
  M 
=   \sqrt{ \sec 2 \phi} 
    \begin{pmatrix}
    \frac{1}{\lambda} \cos ( \theta + \phi ) &  \frac{1}{\lambda} \sin ( \theta + \phi ) \\
    - \lambda \sin (\theta - \phi ) & \lambda \cos ( \theta - \phi )
   \end{pmatrix}
\label{eq:  SL2Rparam} 
\end{equation}
for unique $\lambda$ in the range $0 < \lambda < \infty$.  We will refer to $\theta$ 
as the rotation,
$\phi$ as the obliquity and $\lambda$ as the 
resolution.

The matrix $M$ defines a metric on phase space, with metric tensor $M^{\mathrm{T}} M$:
\begin{equation}
  ds_{M}^{2} 
=  dx_{M}^{2} + dp_{M}^{2} 
= \begin{pmatrix} dx & dp \end{pmatrix} M^{\mathrm{T}} M \begin{pmatrix} dx \\ dp \end{pmatrix}
\label{eq:  Metric}
\end{equation}
($M^{\mathrm{T}}$ being the transpose of $M$).  For a given system some choices
of the matrix $M$, and therefore some choices of metric, will be more natural than others.  
However, 
if one wants to keep the discussion completely general, so that the detailed nature 
of the system is left unspecified, then one must regard the different choices for $M$ as all
being on the same footing---corresponding to the well-known fact, that there is (in general) no 
natural metric on phase space.

From this point of view, the choice of one particular 
conjugate pair $\xOp$ and $\pOp$ as basic, $\xOp_{M}$ and $\pOp_{M}$ being defined
in terms of them, must be regarded as arbitrary.  
Assignments of angle also depend on the choice of metric tensor, 
and must likewise be regarded as arbitrary.  It follows, that in a general context
(though perhaps not in applications to a particular type of system), no 
fundamental significance attaches to the distinction between oblique axes ($\phi \neq 0$) 
and perpendicular axes ($\phi=0$).

It will be shown below that to each $M \in \mathrm{SL}(2,\mathbb{R})$ there corresponds a
retrodictively optimal measurement; and that two such matrices define the same metric if and
only if the corresponding measurements are informationally equivalent.

$M$ has the decomposition
\begin{equation}
  M = T_{\lambda} \, S_{\phi} \, R_{\theta}
\label{eq:  Mdecomp}
\end{equation}
where
\begin{equation} 
  R_{\theta} 
= \begin{pmatrix} \cos \theta & \sin \theta \\ - \sin \theta & \cos \theta \end{pmatrix}   \hspace{0.25 in}
  S_{\phi} 
= \sqrt{\sec 2 \phi} 
  \begin{pmatrix} \cos \phi & \sin  \phi \\ \sin \phi & \cos \phi \end{pmatrix} \hspace{0.25 in}
  T_{\lambda} 
= \begin{pmatrix} \frac{1}{\lambda} & 0 \\ 0& \lambda \end{pmatrix}
\label{eq:  RSTMats}
\end{equation}

Define $\hat{a}= \frac{1}{\sqrt{2}} (\xOp + i \pOp)$.
Let $\hat{U}_{\theta}$ be the unitary rotation operator, 
and $\hat{V}_{\phi}$, $\hat{W}_{\lambda}$
the unitary squeeze operators~\cite{Squeeze} defined by
\begin{equation*}
\begin{split}
      \hat{U}_{\theta} 
  & = \exp \left[ - i \theta \, \hat{a}^{\dagger} \hat{a} \right] \\
      \hat{V}_{\phi} 
  & = \exp \bigl[  \tfrac{i}{2} \tanh^{-1} \left(\tan \phi\right)  
                    \left( \hat{a}^2 + \hat{a}^{\dagger \; 2} \right) 
           \bigr]  \\
      \hat{W}_{\lambda} 
  & = \exp \bigl[ \tfrac{1}{2} \ln \lambda \left(  \hat{a}^2 - \hat{a}^{\dagger \; 2}  \right)\bigr]
\end{split}
\end{equation*}
Then
{\allowdisplaybreaks
\begin{align}
    \hat{U}_{\theta}^{\dagger} 
    \begin{pmatrix}  \xOp \\ \pOp \end{pmatrix}
    \hat{U}_{\theta}^{\vphantom{\dagger}}
& = R_{\theta} \begin{pmatrix}^{\vphantom{\dagger}} \xOp \\ \pOp \end{pmatrix} 
\label{eq:  Umatdef}
\\
    \hat{V}_{\phi}^{\dagger} 
    \begin{pmatrix}  \xOp \\ \pOp \end{pmatrix}
    \hat{V}_{\phi}^{\vphantom{\dagger}}
& = S_{\phi}^{\vphantom{\dagger}} \begin{pmatrix} \xOp \\ \pOp \end{pmatrix} 
\label{eq:  Vmatdef}
\\ 
    \hat{W}_{\lambda}^{\dagger} 
    \begin{pmatrix}  \xOp \\ \pOp \end{pmatrix}
    \hat{W}_{\lambda}^{\vphantom{\dagger}}
& = T_{\lambda}^{\vphantom{\dagger}} \begin{pmatrix} \xOp \\ \pOp \end{pmatrix} 
\label{eq:  Wmatdef}
\end{align}
Hence}
\begin{equation*}
    \begin{pmatrix} \xOp_{M} \\ \pOp_{M} \end{pmatrix}
=   \hat{U}_{\theta}^{\dagger} \hat{V}_{\phi}^{\dagger} \hat{W}_{\lambda}^{\dagger} 
    \begin{pmatrix} \xOp \\ \pOp \end{pmatrix} 
    \hat{W}_{\lambda}^{\vphantom{\dagger}}
    \hat{V}_{\phi}^{\vphantom{\dagger}}
    \hat{U}_{\theta}^{\vphantom{\dagger}} 
\end{equation*}
\section{Informationally Equivalent Measurements}
\label{sec:  meas}
The purpose of this section is to show how the set of retrodictively optimal
measurements divides into subsets of informationally equivalent measurements.

Suppose that we make a retrodictively optimal measurement of the conjugate observables
$\xOp_{M}$, $\pOp_{M}$, of the kind described in ref.~\cite{self2}.  Let $\hat{\mu}_{\mathrm{X}M }$ and
$\hat{\mu}_{\mathrm{P}M }$ be the pointer observables giving the results of the measurements
of $\xOp_{M}$ and $\pOp_{M}$ respectively.  Let $\hat{U}_{\mathrm{meas}}$ be the unitary evolution operator
describing the measurement interaction, and let 
$\hat{\epsilon}_{\mathrm{X}M\mathrm{i}}$, $\hat{\epsilon}_{\mathrm{P} M \mathrm{i}}$ be the retrodictive error operators
\begin{equation}
\begin{split}
     \hat{\epsilon}_{\mathrm{X} M \mathrm{i}} 
 & = \hat{U}_{\mathrm{meas}}^{\dagger} \, 
             \hat{\mu}_{\mathrm{X}M } 
             \hat{U}_{\mathrm{meas}}^{\vphantom{\dagger}} 
      - \xOp_{M} \\
     \hat{\epsilon}_{\mathrm{P} M \mathrm{i}} 
 & = \hat{U}_{\mathrm{meas}}^{\dagger} \, 
             \hat{\mu}_{\mathrm{P}M } 
             \hat{U}_{\mathrm{meas}}^{\vphantom{\dagger}} 
      - \pOp_{M} \\
\end{split}
\label{eq:  RetErrOps}
\end{equation}
as defined in ref.~\cite{self2}.  If the measurement is retrodictively optimal there exists~\cite{self2}
fixed $\tau$ such that
\begin{equation*}
\begin{split}
    \left( \bmat{ \psi \otimes \pt}{\hat{\epsilon}_{\mathrm{X} M \mathrm{i}}^{2}}{ \psi \otimes \pt} \right)^{\frac{1}{2}}
& = \Delta_{\mathrm{ei}} x_{M} =
\frac{\tau}{\sqrt{2}} \\
\left( 
    \bmat{ \psi \otimes \pt
         }{\hat{\epsilon
             }_{\mathrm{P}M\mathrm{i}}^{2}
         }{
           \psi\otimes\pt}
\right)^{\frac{1}{2}} & =
\Delta_{\mathrm{ei}} p_{M} =
\frac{1}{\sqrt{2} \, \tau} \\
\end{split}
\end{equation*}
for every normalised initial system state $\ket{\psi}$ (where $\ket{\pt}$ 
is the initial apparatus state, and 
$\Delta_{\mathrm{ei}} x_{M}$, $\Delta_{\mathrm{ei}} p_{M}$ are the maximal rms errors of
retrodiction~\cite{self1,self2}).

There is no loss of generality in confining 
ourselves to balanced measurements, for which
$\tau = 1$.  In  fact, suppose
that $\tau \neq 1$.  Define
\begin{equation*}
 N = \begin{pmatrix} 
       \frac{1}{\tau} & 0 \\ 0 & \tau
       \end{pmatrix} M
\end{equation*}
It can be seen that measuring the 
observables $\xOp_{M}$, $\pOp_{M}$
to retrodictive accuracies 
$\pm \frac{\tau}{\sqrt{2}}$ and $\pm
\frac{1}{\sqrt{2}\, \tau}$ respectively is
equivalent to measuring the observables
$\xOp_{N}$, $\pOp_{N}$ both to the same
retrodictive accuracy
$\pm \frac{1}{\sqrt{2}}$.

Let $M$ be  the matrix with rotation $\theta$,
obliquity $\phi$ and resolution $\lambda$,
as in Eq.~(\ref{eq:  SL2Rparam}).  Define
\begin{align*}
\xOp_{\theta + \phi} & = \cos(\theta+\phi) \xOp + \sin(\theta+\phi) \pOp 
& \hspace{0.5 in}
\pOp_{\theta - \phi} & = - \sin(\theta-\phi) \xOp + \cos(\theta-\phi) \pOp 
\\ \intertext{Then}
 \xOp_{M} 
&  = \frac{\sqrt{\sec 2 \phi}}{\lambda } \,
     \xOp_{\theta + \phi} 
& \hspace{0.5 in}
     \pOp_{M} 
 & = \lambda \sqrt{\sec 2 \phi} \,
     \pOp_{\theta - \phi}
\end{align*}
It follows, that making a  retrodictively optimal, balanced  measurement
of  $\xOp_{M}$, $\pOp_{M}$ is
equivalent to making a retrodictively optimal measurement of
$\xOp_{\theta + \phi}$,
$\pOp_{\theta - \phi}$ to  accuracies
$\pm \frac{\lambda }{\sqrt{2 \sec 2 \phi}}$
and 
$\pm \frac{ 1}{ \lambda \sqrt{2 \sec 2 \phi} }$
respectively.  This equivalence means that there is associated, to each 
retrodictively optimal measurement of a pair $\xOp_{\theta + \phi}$,
$\pOp_{\theta - \phi}$, a unique matrix $\in \mathrm{SL}(2,\mathbb{R})$.

Let  $M$, $M'$ be any two
matrices $\in \mathrm{SL}(2, \mathbb{R})$, with parameter values
$\theta,\phi,\lambda$ and $\theta',\phi',\lambda'$ respectively.  Suppose that we make a
retrodictively optimal, balanced measurement of the observables 
$\xOp_{M}$, $\pOp_{M}$.  Let $\hat{\mu}_{\mathrm{X}M}$,
$\hat{\mu}_{\mathrm{P}M}$ be the pointer observables representing the result
of this measurement.
Define new pointer
observables $\hat{\mu}_{\mathrm{X}M'}$,
$\hat{\mu}_{\mathrm{P}M'}$:
\begin{equation}
    \begin{pmatrix}
       \hat{\mu}_{\mathrm{X}M'} \\
       \hat{\mu}_{\mathrm{P}M'}
    \end{pmatrix}
=   L
    \begin{pmatrix}
       \hat{\mu}_{\mathrm{X}M} \\
       \hat{\mu}_{\mathrm{P}M}
    \end{pmatrix}
\label{eq:  ptTransform}
\end{equation}
where $L$ is the matrix
\begin{equation*}
  L
=  \begin{pmatrix}
      L_{xx} & L_{xp} \\ L_{px} & L_{pp}
   \end{pmatrix}
= M' M^{-1}
\end{equation*} 
$\hat{\mu}_{\mathrm{X}M'}$,
$\hat{\mu}_{\mathrm{P}M'}$
provide a measurement of $\xOp_{M'}$, $\pOp_{M'}$.
We now ask:  what is the
condition for this measurement  to be 
retrodictively optimal and balanced, the same as the measurement
of $\xOp_{M}$, $\pOp_{M}$?

The retrodictive error operators for the
$M'$-measurement are given by
\begin{equation*}
  \begin{pmatrix}
     \hat{\epsilon}_{\mathrm{X} M' \mathrm{i}} \\
     \hat{\epsilon}_{\mathrm{P} M' \mathrm{i}}
  \end{pmatrix}
= \begin{pmatrix}
     \hat{U}_{\mathrm{meas}}^{\dagger} \, 
        \hat{\mu}_{\mathrm{X}M' } 
     \hat{U}_{\mathrm{meas}}^{\vphantom{\dagger}} 
      - \xOp_{M'} 
  \\
     \hat{U}_{\mathrm{meas}}^{\dagger} \, 
             \hat{\mu}_{\mathrm{P}M' } 
     \hat{U}_{\mathrm{meas}}^{\vphantom{\dagger}} 
      - \pOp_{M'} 
   \end{pmatrix}
=  L
  \begin{pmatrix}
     \hat{\epsilon}_{\mathrm{X} M \mathrm{i}} \\
     \hat{\epsilon}_{\mathrm{P} M \mathrm{i}}
  \end{pmatrix}
\end{equation*}
[see Eq.~(\ref{eq:  RetErrOps})].
Therefore
\begin{equation*}
\begin{split}
     \bigl<
       \hat{\epsilon}_{ \mathrm{X}
                        M' \mathrm{i}}^{\; 2}
     \bigr> 
& =  L_{xx}^{2} 
     \bigl< 
       \hat{\epsilon}_{\mathrm{X}
                        M\mathrm{i}}^{\; 2}
     \bigr> 
     +
     L_{xp}^2
     \bigl< 
       \hat{\epsilon}_{\mathrm{P} M\mathrm{i}}^{\; 2}
     \bigr>
     +
     L_{xx} L_{xp}
     \bigl< (
       \hat{\epsilon}_{\mathrm{X} M\mathrm{i}}
       \hat{\epsilon}_{\mathrm{P} M\mathrm{i}}
       +
       \hat{\epsilon}_{\mathrm{P} M\mathrm{i}}
       \hat{\epsilon}_{\mathrm{X} M\mathrm{i}})
     \bigr>
\\
     \bigl<
       \hat{\epsilon}_{ \mathrm{X}
                        M' \mathrm{i}}^{\; 2}
     \bigr> 
& =  L_{px}^{2} 
     \bigl< 
       \hat{\epsilon}_{\mathrm{X}
                        M\mathrm{i}}^{\; 2}
     \bigr> 
     +
     L_{pp}^2
     \bigl< 
       \hat{\epsilon}_{\mathrm{P} M\mathrm{i}}^{\; 2}
     \bigr>
     +
     L_{px} L_{pp}
     \bigl< (
       \hat{\epsilon}_{\mathrm{X} M\mathrm{i}}
       \hat{\epsilon}_{\mathrm{P} M\mathrm{i}}
       +
       \hat{\epsilon}_{\mathrm{P} M\mathrm{i}}
       \hat{\epsilon}_{\mathrm{X} M\mathrm{i}})
     \bigr>
\end{split}
\end{equation*}
The fact that the $M$-measurement is retrodictively
optimal and balanced means that 
$\bigl< 
 \hat{\epsilon}_{\mathrm{X}
 M\mathrm{i}}^{\; 2}
 \bigr>
 =
 \bigl< 
 \hat{\epsilon}_{\mathrm{P} M\mathrm{i}}^{\; 2}
 \bigr>
 = \frac{1}{2}$.
Also, we have from Lemma 2, proved in 
ref.~\cite{self2}, that
\begin{equation*}
  \left(
  \hat{\epsilon}_{\mathrm{X} M\mathrm{i}}
  + i \hat{\epsilon}_{\mathrm{P} M\mathrm{i}}
  \right)
  \ket{\psi \otimes \pt}
= 0
\end{equation*}
for every initial system state $\ket{\psi}$ (where 
$\ket{\pt}$ is the initial apparatus state,
as before).  Hence
\begin{equation*}
     \left< \left(
            \hat{\epsilon}_{\mathrm{X} M\mathrm{i}}
            \hat{\epsilon}_{\mathrm{P} M\mathrm{i}}
            +
            \hat{\epsilon}_{\mathrm{P} M\mathrm{i}}
             \hat{\epsilon}_{\mathrm{X} M\mathrm{i}}
     \right)^{\vphantom{2}} \right>
  =   - i \left<
          \left(
             \hat{\epsilon}_{\mathrm{X} M\mathrm{i}}
          + i \hat{\epsilon}_{\mathrm{P} M\mathrm{i}}
            \right)^2
         \right>
 = 0
\end{equation*}
Consequently
\begin{equation*}
\begin{split}
     \bigl<
       \hat{\epsilon}_{ \mathrm{X}
                        M' \mathrm{i}}^{\; 2}
     \bigr> 
& =  \frac{1}{2} 
     \left( L_{xx}^{2} + L_{xp}^2 \right)
\\
     \bigl<
       \hat{\epsilon}_{ \mathrm{X}
                        M' \mathrm{i}}^{\; 2}
     \bigr> 
& =  \frac{1}{2} 
     \left(  L_{px}^{2}    +  L_{pp}^2 \right)
\end{split}
\end{equation*}
It follows that the 
$M'$-measurement is retrodictively optimal and
balanced if and only if 
\begin{equation*}
   L_{xx}^{2} + L_{xp}^2 
=  L_{px}^{2}    +  L_{pp}^2
= 1
\end{equation*}
which implies
\begin{equation*}
  L L^{\mathrm{T}} 
= \begin{pmatrix}
    1 & L_{xx} L_{px} + L_{xp} L_{pp} \\
    L_{xx} L_{px} + L_{xp} L_{pp} & 1
  \end{pmatrix}
\end{equation*}
where $L^{\mathrm{T}}$ denotes the transpose
of $L$.
Since $\Det L = 1$ we must have
$L_{xx} L_{px} + L_{xp} L_{pp}=0$, which means that
$L$ is a rotation matrix.  We conclude,
that the necessary and sufficient condition for
the $M'$-measurement to be retrodictively optimal
and balanced is that
\begin{equation*}
  M' = R_{\psi} M
\end{equation*}
for some $\psi$ [where $R_{\psi}$ is a rotation matrix as defined in Eq.~(\ref{eq: 
RSTMats})].   If $M$ and $M'$ satisfy this condition we will say that the corresponding
measurements are informationally equivalent, and we will write $M \sim M'$.

If $M \sim M'$ it means that a retrodictively optimal
measurement of
$x_{\theta+\phi}$,
$p_{\theta-\phi}$ to  accuracies $\pm \frac{\lambda}{\sqrt{2 \sec{2\phi}} }$,
$\pm \frac{1}{ \lambda \, \sqrt{2 \sec{2 \phi}} }$ yields exactly the same information as a
retrodictively optimal measurement of $x_{\theta'+\phi'}$,
$p_{\theta'-\phi'}$ to  accuracies $\pm \frac{\lambda'}{\sqrt{2 \sec{2\phi'}} }$,
$\pm \frac{1}{ \lambda' \, \sqrt{2 \sec{2 \phi'}} }$

The condition for $M \sim M'$ can alternatively be written
\begin{equation}
{M'}^{\mathrm{T}} M' = M{\vphantom{'}}^{\mathrm{T}} M
\label{eq:  EquivCondB}
\end{equation}
In other words, two measurements are informationally equivalent if and only if the
corresponding matrices define the same phase-space metric [see Eq.~(\ref{eq:  Metric})].

We conclude this section by showing, that given any 
$M \in \mathrm{SL}(2,\mathbb{R})$, it is possible to find $M_{0} \sim M$ with 
zero obliquity (so that the axes are perpendicular).

We can write
\begin{equation}
  M^{\mathrm{T}} M
= \begin{pmatrix} a &  c \\ 
  c & b \end{pmatrix}
\label{eq:  abcDef}
\end{equation}
where $a$, $b > 0$  and
$ab - c^2 =1$.   Let $M_{0}$ be a matrix with
zero obliquity:
\begin{equation*}
  M_{0}
= \begin{pmatrix}
     \frac{1}{\lambda_{0}} \cos \theta_{0}
  &   \frac{1}{\lambda_{0}} \sin \theta_{0}
  \\  - \lambda_{0} \sin \theta_{0}
  &   \lambda_{0} \cos \theta_{0}
  \end{pmatrix}
\end{equation*}
[see Eq.~(\ref{eq:  SL2Rparam})].  We want to
show that it is possible to choose
$\lambda_{0}$, $\theta_{0}$ so that
\begin{equation*}
  M_{0}^{\mathrm{T}} M_{0} 
= M_{\vphantom{0}}^{\mathrm{T}} M
\end{equation*}
which means
\begin{equation*}
\begin{split}
   a - b
&  =  \left(  \frac{1}{\lambda_{0}^{2}}
           - \lambda_{0}^{2}
       \right) \cos 2 \theta_{0}
\\ a + b
&  =  \left(  \frac{1}{\lambda_{0}^{2}}
           + \lambda_{0}^{2}
       \right)
\\ 2 c
&  =  \left(  \frac{1}{\lambda_{0}^{2}}
           - \lambda_{0}^{2}
       \right) \sin 2 \theta_{0}
\end{split}
\end{equation*}
It is readily confirmed that these equations
are  soluble.  An explicit solution is
\begin{align}
  \theta_{0}
& = \begin{cases}
     \frac{1}{2}
     \tan^{-1} \frac{2 c}{b - a}
     \hspace{0.5 in} &
     \text{if}\ b \neq a
  \\ - \frac{\pi}{4}
     \hspace{0.5 in} & 
     \text{if}\ b = a
   \end{cases}
\label{eq:  thet0}
\\
\intertext{and}
   \lambda_{0}
& =  \begin{cases}
     \frac{1}{\sqrt{2}}
     \left( (a+b)
           -  \sign (a-b)
            \left( (a+b)^2 - 4\right)^{\frac{1}{2}}
      \right)^{\frac{1}{2}}
      \hspace{0.5 in} &
      \text{if}\ b \neq a
     \\
      \sqrt{a+c}
      \hspace{0.5 in} &
      \text{if}\ b = a
    \end{cases}
\label{eq:  lam0}
\end{align}
Using these formulae, it is straightforward to express 
$\theta_{0}$, $\lambda_{0}$ directly in terms of
the parameters
$\theta$, $\phi$, $\lambda$.  For the sake of simplicity we  confine ourselves to the case
$\lambda=1$, when
\begin{align*}
  \theta_{0}
& = \begin{cases} 
      [\theta]_{\frac{\pi}{2}} - \frac{\pi}{4} \hspace{0.5 in} 
   &  \text{if $ \phi \neq 0$} \\
      - \frac{\pi}{4} \hspace{0.5 in}
   &  \text{if $\phi =0$}
   \end{cases}
\\
\intertext{and}
   \lambda_{0}
& = \begin{cases}
      \bigl( \sec 2 \phi + \sign (\sin 2 \theta)
\tan 2 \phi\bigr)^{\frac{1}{2}}
      \hspace{0.5 in} & \text{if $\sin 2 \theta  \neq 0$}
    \\
      \bigl( \sec 2 \phi + \cos 2 \theta \tan 2
\phi\bigr)^{\frac{1}{2}}
      \hspace{0.5 in} & \text{if $\sin 2 \theta  = 0$}
    \end{cases}
\end{align*}
where the notation $[\theta]_{\frac{\pi}{2}}$ 
means ``$\theta$ mod $\frac{\pi}{2}$'':
\begin{equation*}
[\theta]_{\frac{\pi}{2}} =
\theta - \frac{n \pi}{2}
\hspace{0.5 in}
\text{if $\frac{n \pi}{2} \le \theta <
\frac{(n+1)\pi}{2}$}
\end{equation*}
for every integer $n$.

\begin{figure}
\includegraphics{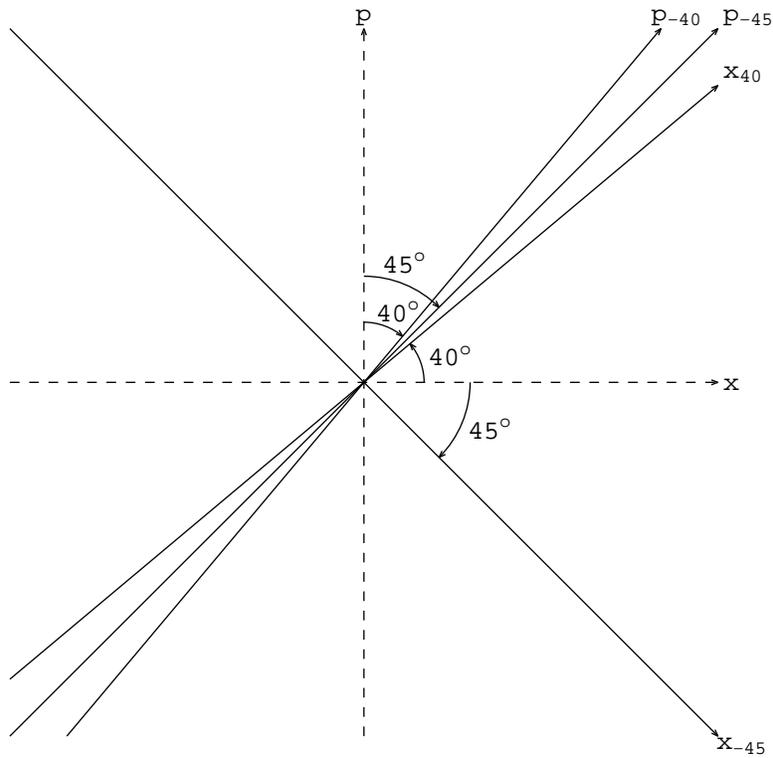}
\caption{Equivalence of oblique and orthogonal phase space coordinate systems.  Measuring the
oblique coordinates
$x_{40}$ and 
$p_{-40}$ each to the same accuracy $\pm 
\frac{1}{\sqrt{2 \sec 80^o}} = \pm 0.29$
is equivalent to measuring
the orthogonal  coordinates $x_{- 45}$ and
$p_{- 45}$ to accuracies $\pm 
\frac{\lambda_0}{\sqrt{2}} = \pm 2.39$
 and $\pm \frac{1}{\sqrt{2} \lambda_0} = \pm
0.21$
respectively. The measurement of $p_{- 45}$
is more accurate than the measurements of 
$x_{40}$ and $p_{-40}$.  The measurement 
of  $x_{- 45}$ is much less accurate.}
\label{fig:  pic2}
\end{figure}

This means that a
retrodictively optimal measurement of 
the non-orthogonal quadratures
$\hat{x}_{\theta + \phi}$, 
$\hat{p}_{\theta - \phi}$ 
to accuracies $\pm \frac{\lambda}{\sqrt{2 \sec 2
\phi}}$ and
$\pm \frac{1}{\lambda \sqrt{ 2 \sec 2 \phi}} $
yields  the same  information as
a measurement of the
orthogonal quadratures $\hat{x}_{\theta_{0}}$, 
$\hat{p}_{\theta_{0}}$ to accuracies
$\pm \frac{\lambda_{0}}{\sqrt{2}}$ and
$\pm \frac{1}{\sqrt{2} \lambda_{0}}$.
Fig.~\ref{fig:  pic2} gives an illustration, for the case
$\lambda=1$, $\theta = 0$, $\phi =
40^{\mathrm{o}}$ [implying 
$\theta_0 = - 45^{\mathrm{o}}$ and
$\lambda_0 = \left(\sec 80^{\mathrm{o}}
+\tan
80^{\mathrm{o}}\right)^{\frac{1}{2}}$].
\section{The Distribution of Measured
Values}
\label{sec:  distribution}
Let $M$ be any matrix $\in \mathrm{SL} (2,
\mathbb{R})$, and consider a balanced retrodictively
optimal measurement of the observables
$\xOp_{M}$, $\pOp_{M}$.  Let
$\hat{\mu}_{\mathrm{X}M }$ and
$\hat{\mu}_{\mathrm{P}M }$ be the pointer
observables describing the outcome of this
measurement. Define $\MxOp$,
$\MpOp$ by
\begin{equation*}
  \begin{pmatrix}
  \MxOp \\ \MpOp
  \end{pmatrix}
=  M^{-1}
   \begin{pmatrix}
   \hat{\mu}_{\mathrm{X}M } \\
   \hat{\mu}_{\mathrm{P}M }
   \end{pmatrix}
\end{equation*}
(\emph{c.f.} Eq.~(\ref{eq:  ptTransform})). 
Then $\MxOp$, $\MpOp$ are the pointer
observables for a measurement of $\xOp$ and
$\pOp$.  This measurement will not be
retrodictively optimal and balanced unless 
$M^{\mathrm{T}} M = 1$.

Let $\rho_{M}$ be the probability
density function describing the result of the
measurement of 
$\xOp_{M}$, $\pOp_{M}$; and let 
$Q_{M}$ be the probability
density function describing the result of the
measurement  of $\xOp$, $\pOp$.  Then
\begin{equation*}
   Q_{M} (x,p) =
\rho_{M}^{\vphantom{\dagger}}
(x_M^{\vphantom{\dagger}},p_M^{\vphantom{\dagger}})
\end{equation*}
for all $x$, $p$, where
\begin{equation*}
  \begin{pmatrix}
  x_M^{\vphantom{\dagger}} \\
p_M^{\vphantom{\dagger}}
  \end{pmatrix}
=  M
   \begin{pmatrix}
   x \\
   p
   \end{pmatrix}
\end{equation*}

Let $\hat{a}_{M}^{\vphantom{\dagger}}$ and 
 $\ket{(x,p)_M}$ be the annihilation operator and (normalised)
squeezed state defined by
\begin{align}
   \hat{a}_{M}^{\vphantom{\dagger}} 
& =  \frac{1}{\sqrt{2}}
   \left(  \xOp_{M} + i \pOp_{M}
   \right)  
\notag 
\\
\hat{a}_M 
   \bket{(x,p)_M^{\vphantom{\dagger}}}
& =  \frac{1}{\sqrt{2}}
   \left( x_M + i \, p_{M} \right)
   \bket{(x,p)_M^{\vphantom{\dagger}}} 
\label{eq:  xpMketdef}
\end{align}
Using the result proved in 
ref.~\cite{self2} we have
\begin{equation}
   Q_{M}^{\vphantom{\dagger}} (x,p)
=  \rho_{M}^{\vphantom{\dagger}} 
   (x_M^{\vphantom{\dagger}},
   p_M^{\vphantom{\dagger}})
=  \frac{1}{ 2 \pi}
   \bmat{ (x,p)_M^{\vphantom{\dagger}}
       }{ \hat{\rho}
       }{ (x,p)_M^{\vphantom{\dagger}}
       }
\label{eq:  QMdef} 
\end{equation}
where $\hat{\rho}$ is the density matrix
describing the initial state of the system.  We see from this
that $Q_{M}$ is a generalised Husimi function of the kind defined 
in Section~\ref{sec:  intro}.

It was shown in the last
section that there exists $\psi$ such that
\begin{equation*}
   M = R_{\psi} T_{\lambda_0} R_{\theta_0}
\end{equation*}
where $\lambda_0$, $\theta_0$ are the
quantities defined by Eqs.~(\ref{eq:  thet0})
and~(\ref{eq:  lam0}),
and where $R_{\psi}$, $T_{\lambda_0}$,
$R_{\theta_0}$ are the matrices defined 
by Eq.~(\ref{eq:  RSTMats}).
In view of 
Eqs.~(\ref{eq:  Umatdef}-\ref{eq:  Wmatdef})
we then have~\cite{Squeeze}
\begin{equation}
  \bket{(x,p)_{M}^{\vphantom{dagger}}}
= e^{i \chi} \hat{D}_{xp} \hat{U}_{\theta_0}^{\dagger}
   \hat{W}_{\lambda_0}^{\dagger} 
   \hat{U}_{\psi}^{\dagger} \bket{0}
=  e^{i \chi} \hat{D}_{xp} \hat{U}_{\theta_0}^{\dagger}
   \hat{W}_{\lambda_0}^{\dagger} \bket{0}
\label{eq:  xpMtermsD} 
\end{equation}
where $\chi$ is a phase, $\hat{D}_{xp}$ is the displacement
operator
\begin{equation*}
   \hat{D}_{xp}
=  \exp \left[ - i(x_{M} \, \pOp_{M} - p_{M}\,
   \xOp_{M})\right]
=   \exp \left[ - i(x \pOp - p \xOp)\right]
\end{equation*}
and where $\ket{0}$ is the vacuum state annihilated by
$\hat{a} = \frac{1}{\sqrt{2}} (\xOp + i \pOp)$.
Hence
\begin{align}
& Q_{M} (x,p)
\notag
\\
& \hspace{0.25 in}
= \frac{1}{\pi}
  \int dx' dp' \,
  \exp 
   \left[ - a (x'-x)^2 
          - 2 c (x'-x)(p'-p)
          - b (p'-p)^2
   \right]
   W(x',p')
\label{eq:  QMtermsWig1}
\end{align}
where $a$, $b$, $c$ are the quantities defined 
in Eq.~(\ref{eq:  abcDef}) and $W$ is the Wigner function describing
the initial system state. 
It can be seen that the quadratic form occurring in the argument of the exponential is the
phase space metric corresponding to $M$ [\emph{c.f.}\ Eqs.~(\ref{eq:  Metric}) and~(\ref{eq: 
abcDef})].  This means that two retrodictively optimal measurements define the same
distribution if and only if they are informationally equivalent---as was to be expected.

In  terms of the parameters $\lambda_0$, $\theta_0$ we have
\begin{equation*}
  Q_{M} (x, p)
= \frac{1}{\pi}
  \int dx' dp' \,
    \exp\left[- \frac{1}{\lambda_0^2} 
                 (x'_{\theta_0} - x^{\vphantom{t}}_{\theta_0})^2 
              - \lambda_0^2 (p'_{\theta_0} - p^{\vphantom{'}}_{\theta_0})^2
        \right] W(x',p')
\end{equation*}
It can be seen that $Q_{M}$ is obtained by smoothing the Wigner function
on the scale $\lambda_0$ parallel to the $x_{\theta_0}$ axis, and on the scale
$\frac{1}{\lambda_0}$ parallel to the $p_{\theta_0}$ axis.

\newpage
\newpage

\end{document}